# Unified model of voltage/current mode control to predict saddle-node bifurcation

Chung-Chieh Fang *

## SUMMARY

A unified model of voltage mode control (VMC) and current mode control (CMC) is proposed to predict the saddle-node bifurcation (SNB). Exact SNB boundary conditions are derived, and can be further simplified in various forms for design purpose. Many approaches, including steady-state, sampled-data, average, harmonic balance, and loop gain analyses are applied to predict SNB. Each approach has its own merits and complement the other approaches.

**KEY WORDS:** DC-DC power conversion, voltage mode control, current mode control, saddle-node bifurcation, sampled-data analysis, harmonic balance analysis

---

*C.-C Fang is with Advanced Analog Technology, 2F, No. 17, Industry E. 2nd Rd., Hsinchu 300, Taiwan, Tel: +886-3-5633125 ext 3612, Email: fangcc3@yahoo.com

# Contents

# 1 Introduction

Occurrence of saddle-node bifurcation (SNB) [1] is generally unnoticed. A "different" instability and a possibility of two steady-state solutions in the buck converter under current mode control (CMC) in discontinuous conduction mode (DCM) have been reported [2, 3]. The "different" instability is actually a saddle-node bifurcation. In some specific instances when SNB is explicitly noticed [4, 5], no SNB boundary conditions have been derived.

In this paper, a unified model of voltage mode control (VMC) and CMC is proposed to predict SNB. Based on the unified model, seven different approaches are applied to analyze SNB. Exact boundary conditions are derived based on these approaches. The boundary conditions define the critical values of converter parameters when SNB occurs. Applying different approaches actually leads to the same boundary conditions, further confirming the accuracy of the derived boundary conditions. The exact boundary conditions can be further simplified in various approximate closed forms for design purpose.

The SNB may explain some sudden *jumps* or *disappearances* of steady-state solutions observed in DC-DC switching converters. Suppose SNB occurs at a critical parameter value. On one side of the critical parameter value (also the bifurcation point), there are multiple solutions (attractors) with different domains of attraction. When the system is perturbed, the state of the converter may *jump* from one attractor to another. On the other side of the critical parameter value, no solution exists. When the parameter crosses the critical value, the converters originally have multiple solutions, but now these solutions suddenly *disappear*. Those phenomena such as multiple solutions, sudden jumps and disappearances are undesirable in DC-DC switching converters. It is useful to understand the dynamics of SNB and its boundary condition to avoid its occurrence.

The SNB actually exists in popular DC-DC converters, such as buck or boost converters, under various control schemes, such as VMC, CMC, or multi-loop state feedback control. The seven different approaches are applied to accurately predict occurrence of SNB in buck and boost converters under these control schemes. Without loss generality, only continuous conduction mode (CCM) is considered. Analysis of SNB in DCM is reported separately [6].

The remainder of the paper is organized as follows. In Section 2, the operation of VMC and CMC is modeled in a single unified block diagram. In Section 3, steady-state analysis and small-signal analysis are presented. In Section 4, based on sampled-data slope-based analysis, general SNB boundary conditions for buck and boost converters are derived. In Sections 5-7, average, harmonic balance, and loop gain analyses are applied to predict SNB. In Sections 8 and 9, various approaches are applied to analyze SNB in buck and boost converters, respectively. Conclusions are collected in Section 10.

# 2 Operation of Voltage/Current Mode Control

The operation of a DC-DC switching converter under VMC or CMC can be described *exactly* by a unified block diagram model developed in [7, 8] shown in Fig. 1. The control (reference) signal $v_r$ controls the output voltage $v_o$ in VMC. In CMC, $v_r$ is generally denoted as $i_c$ to control the (peak) inductor current $i_L$. Denote the source voltage as $v_s$. In the model, $A_1, A_2 \in \mathbf{R}^{N \times N}$, $B_1, B_2 \in \mathbf{R}^{N \times 2}$, $C, E_1, E_2 \in \mathbf{R}^{1 \times N}$, and $D \in \mathbf{R}^{1 \times 2}$ are constant matrices, where $N$ is the system dimension. For example, $N = 5$ for a buck converter with a type III compensator. Within a clock period $T$, the dynamics is switched between two stages, $S_1$ and $S_2$. Switching occurs when the ramp signal $h(t)$ intersects with the compensator output $y := Cx + Du \in \mathbf{R}$. Denote the ramp amplitude as $V_h$, and denote the switching frequency as $f_s := 1/T$ and let $\omega_s := 2\pi f_s$.

Typical signal waveforms for VMC and CMC are shown in Fig. 2 and Fig. 3, respectively. In

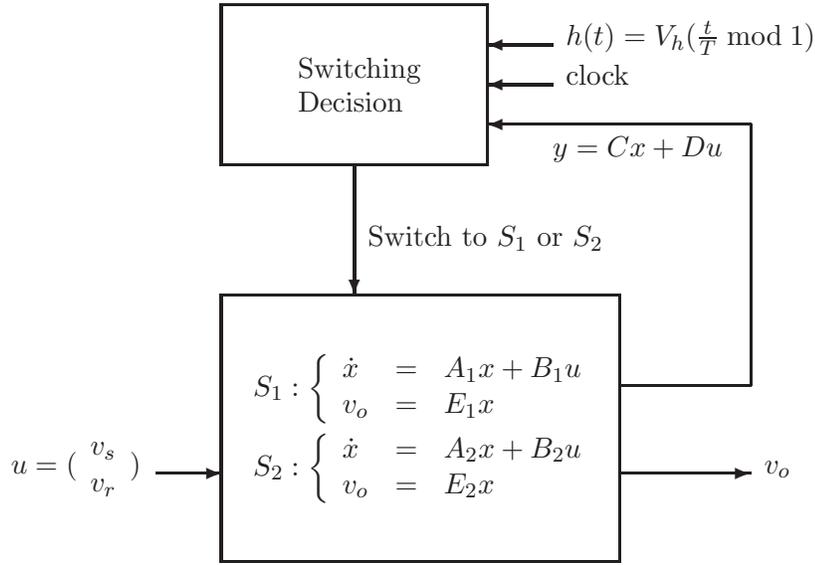

Figure 1: Block diagram model for switching converter

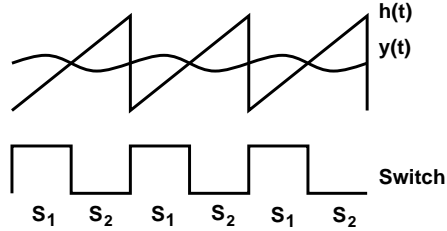

Figure 2: Waveforms for voltage mode control

Fig. 3, the ramp has positive slope, instead of negative slope as commonly seen in most literature, in order to be consistent with VMC. Other control schemes (average current mode control, for example) also fit the model of Fig. 1.

## 3 Steady-State and Small-Signal Analysis

### 3.1 Periodic Orbits as Steady-State Solutions

The periodic solution $x^0(t)$ of the system in Fig. 1 corresponds to a fixed point $x^0(0)$ in the sampled-data dynamics. A typical periodic solution $x^0(t)$ is shown in Fig. 4, where $\dot{x}^0(d^-) = A_1 x^0(d) + B_1 u$ and $\dot{x}^0(d^+) = A_2 x^0(d) + B_2 u$ denote the time derivative of $x^0(t)$ at $t = d^-$ and $d^+$, respectively. Let $y^0(t) = Cx^0(t) + Du$. In steady state, $\dot{y}^0(t) = C\dot{x}^0(t)$. Let the steady-state duty cycle be $D$ and $d := DT$. Confusion of notations for capacitance $C$ and duty cycle $D$ with the matrices $C$ and $D$ can be avoided from the context.

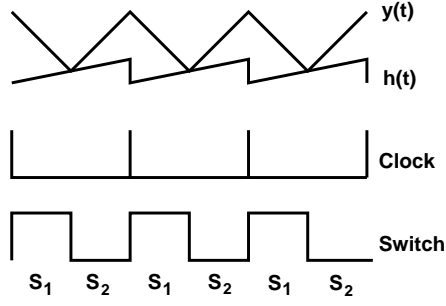

Figure 3: Waveforms for current mode control

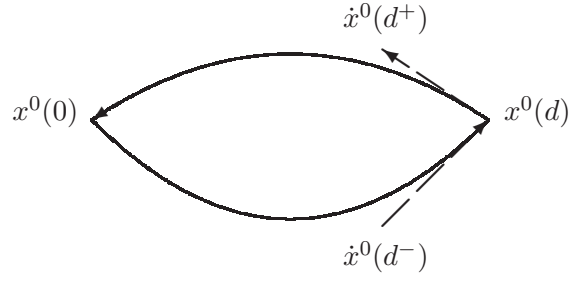

Figure 4: A typical periodic solution $x^0(t)$ of a DC-DC converter in state space

In steady state,

$$x^0(d) = e^{A_1 d} x^0(0) + \int_0^d e^{A_1 \sigma} d\sigma B_1 u \tag{1}$$

$$x^0(0) = e^{A_2(T-d)} x^0(d) + \int_0^{T-d} e^{A_2 \sigma} d\sigma B_2 u \tag{2}$$

From (1) and (2), one has

$$x^0(d) = (I - e^{A_1 d} e^{A_2(T-d)})^{-1} (e^{A_1 d} \int_0^{T-d} e^{A_2 \sigma} d\sigma B_2 u + \int_0^d e^{A_1 \sigma} d\sigma B_1 u) \tag{3}$$

Let $B_1 := [B_{11}, B_{12}]$, $B_2 := [B_{21}, B_{22}]$ to expand the matrices into two columns. The buck converter generally has $A_1 = A_2$ (invertible), $B_{21} = 0$, and $B_{12} = B_{22}$. Then, from (1) and (2), one has

$$x^0(d) = (I - e^{A_1 T})^{-1} A_1^{-1} (e^{A_1 d} - I) B_{11} v_s - A_1^{-1} B_{12} v_r \tag{4}$$

The boost converter generally has $B_1 = B_2$, then

$$x^0(d) = (I - e^{A_1 d} e^{A_2(T-d)})^{-1} (e^{A_1 d} \int_0^{T-d} e^{A_2 \sigma} d\sigma + \int_0^d e^{A_1 \sigma} d\sigma) B_1 u := X(d) B_1 u \tag{5}$$

## 3.2 Boundary Conditions Derived From Steady-State Analysis

In steady state,
$$Cx^0(d) + Du - h(d) = 0 \qquad (6)$$
which is an equation in terms of $d = DT$. Generally, if this equation has multiple solutions of $d$, SNB may occur if a converter parameter varies. When SNB occurs, only a *single* solution exists, which means that the curve $Cx^0(d) + Du - h(d)$ has a flat slope when SNB occurs.

Take the buck converter for example, using (4), and differentiate (6) with respective to $d$, one has the SNB boundary condition,
$$TC(I - e^{A_1 T})^{-1} e^{A_1 d} B_{11} v_s - V_h = 0 \qquad (7)$$
It will be shown later that applying different approaches also leads to the same condition.

## 3.3 Small-Signal Analysis

Using a hat ˆ to denote small perturbations (e.g., $\hat{x}_n = x_n - x^0(0)$). From [7, 8, 9], the linearized sampled-data dynamics is
$$\hat{x}_{n+1} = \Phi \hat{x}_n \qquad (8)$$
where the Jacobian matrix [8]
$$\Phi = e^{A_2(T-d)}(I - \frac{(\dot{x}^0(d^-) - \dot{x}^0(d^+))C}{\dot{y}^0(d^-) - \dot{h}(d)}) e^{A_1 d} \qquad (9)$$

SNB occurs when one eigenvalue of $\Phi$ is 1, and $\det[I - \Phi] = 0$. Since other bifurcations [10], such as pitchfork or transcritical bifurcations, also have one eigenvalue of $\Phi$ at 1, the same analysis can be applied to analyze these bifurcations, omitted to save space. Although the *general methodology* of sampled-data analysis has been known in the last three decades, the *closed form* expression of (9) was first derived and published in [7, 9], to the author's knowledge. All slope-based boundary conditions derived in this paper are based on the closed form expression of (9).

# 4 Sampled-Data Slope-Based Analysis

## 4.1 General Boundary Conditions

Based on $\det[I - \Phi] = 0$, the following theorem is obtained [7, p. 46].

**Theorem 1** *In a system as shown in Fig. 1, saddle-node bifurcation occurs when*
$$\dot{y}^0(d^-) - C(I - e^{-A_2(T-d)} e^{-A_1 d})^{-1}(\dot{x}^0(d^-) - \dot{x}^0(d^+)) = \dot{h}(d) \qquad (10)$$

The proof is as follows. Suppose 1 is not an eigenvalue of $e^{A_2(T-d)} e^{A_1 d}$, then

$$\det[I - \Phi] = \det[I - e^{A_2(T-d)} e^{A_1 d}] \det[I + (I - e^{A_2(T-d)} e^{A_1 d})^{-1} e^{A_2(T-d)} \frac{\dot{x}^0(d^-) - \dot{x}^0(d^+)}{C\dot{x}^0(d^-) - \dot{h}(d)} C e^{A_1 d}]$$

$$= \det[I - e^{A_2(T-d)} e^{A_1 d}][1 + C e^{A_1 d}(I - e^{A_2(T-d)} e^{A_1 d})^{-1} e^{A_2(T-d)} \frac{\dot{x}^0(d^-) - \dot{x}^0(d^+)}{\dot{y}^0(d^-) - \dot{h}(d)}]$$

$\det[I - \Phi] = 0$ requires that the last term (inside of the second square brackets) of the last equation equals to zero, which leads to (10).

One can expand (10) in terms of $x^0(d)$,

$$C(A_1 x^0(d) + B_1 u) - C(I - e^{-A_2(T-d)} e^{-A_1 d})^{-1}((A_1 - A_2) x^0(d) + (B_1 - B_2) u) = \dot{h}(d) \quad (11)$$

Note that the condition (10) is valid for either VMC and CMC, and applicable to *general* switching converters of any system dimension. Also note that in (10), the left side is in terms of the ripple slopes ($\dot{y}^0(d^-)$, $\dot{x}^0(d^-)$, and $\dot{x}^0(d^+)$), and the right side is the ramp slope $\dot{h}(d)$. As in the popular slope-based boundary condition for subharmonic oscillation in CMC, the condition (10) is also slope-based.

The condition (10) can be proved to be equivalent to

$$\dot{y}^0(d^+) - C(e^{A_2(T-d)} e^{A_1 d} - I)^{-1}(\dot{x}^0(d^-) - \dot{x}^0(d^+)) = \dot{h}(d) \quad (12)$$

Using which one of (10) or (12) depends on convenience. Since the proof for (10) is given, the proof for (12) is omitted to save space.

### 4.2 Buck Converter

The buck converter generally has $A_1 = A_2$, $B_{21} = 0$, and $B_{12} = B_{22}$. Using (4), the boundary condition (11) becomes

$$C(I - e^{A_1 T})^{-1} e^{A_1 d} B_{11} v_s = \dot{h}(d) \quad (13)$$

or in terms of $v_s$, which shows the critical value of $v_s$ when SNB occurs,

$$v_s = \frac{\dot{h}(d)}{C(I - e^{A_1 T})^{-1} e^{A_1 d} B_{11}} \quad (14)$$

Both (13) and (14) are equivalent to (7) which is based on the steady-state analysis.

Based on the assumption that $RC$ and $\sqrt{LC}$ are much larger than $T$, matrix approximations such as $e^{A_1 T} \approx I + A_1 T + A_1^2 T^2/2$ and $(I + A_1 T)^{-1} \approx I - A_1 T$ can be applied. Then, the boundary condition (13) leads to

$$-\frac{1}{T} C A^{-1} B_{11} + (\frac{1}{2} - D) C B_{11} - (\frac{1 - 6D + 6D^2}{12}) C A_1 B_{11} T \approx \frac{\dot{h}(d)}{v_s} \quad (15)$$

or in terms of $v_s$,

$$v_s \approx \frac{\dot{h}(d)}{-\frac{1}{T} C A^{-1} B_{11} + (\frac{1}{2} - D) C B_{11} - (\frac{1 - 6D + 6D^2}{12}) C A_1 B_{11} T} \quad (16)$$

*Remarks:*

(a) The left side of the boundary condition (15) is a weighted combination of $CA^{-1} B_{11}/T$, $CB_{11}$ and $CA_1 B_{11} T$. It will be shown that if the equivalent series resistance (ESR) $R_c = 0$, $CA_1 B_{11}$ dominates in VMC, while $CB_{11}$ dominates in CMC. For $R_c > 0$, either VMC or CMC has both the terms $CB_{11}$ and $CA_1 B_{11} T$, indicating that the SNB conditions for VMC and CMC are closely related.

(b) The boundary condition (15) seems to have a pattern. It is a hypothesis that the *exact* boundary condition has the form

$$C(\sum_{n=-1}^{\infty} \delta_n(D) A_1^n T^n) B_{11} = \frac{\dot{h}(d)}{v_s} \quad (17)$$

where $\delta_{-1}(D) = -1$, $\delta_0(D) = (1-2D)/2$, $\delta_1(D) = (-1+6D-6D^2)/12$, etc., and $d\delta_{n+1}(D)/dD = \delta_n(D)$. Further research on the series expression of (17) will be reported separately. However, since an exact condition as in (13) has been obtained, another exact condition in series expression may give additional insights but may be unnecessary.

(c) The condition (13) is exact, while (15) is approximate. Based on simulations, if the poles are smaller than $\omega_s/10$, the approximate condition (15) is close to the exact condition (13), and it is generally adequate to predict SNB. For poles greater than $\omega_s/10$, using the exact condition (13) gives more accurate results.

(d) When the switching frequency is high ($T$ is small), then the first term of (15) dominates, and the boundary condition (15) becomes

$$-\frac{1}{T}CA^{-1}B_{11} \approx \frac{\dot{h}(d)}{v_s} \tag{18}$$

It will be shown later that this is the boundary condition derived from the average analysis.

### 4.3 Boost Converter

Let $\Lambda(d) := I + (A_1 - (I - e^{-A_2(T-d)}e^{-A_1d})^{-1}(A_1 - A_2))X(d)$ to simplify the equation. The analysis for the boost converter is similar to that for the buck converter. Using (5), the boundary condition (11) becomes

$$C\Lambda(d)B_1 u = \dot{h}(d) \tag{19}$$

or in terms of $v_s$,

$$v_s = \frac{\dot{h}(d) - C\Lambda(d)B_{12}v_r}{C\Lambda(d)B_{11}} \tag{20}$$

### 4.4 The "S plot": a Slope-Based Plot in the Real Domain

Define an "S plot" as a function of $D = d/T$,

$$\begin{align}
S(D) &:= \dot{y}^0(d^-) - C(I - e^{-A_2(T-d)}e^{-A_1d})^{-1}(\dot{x}^0(d^-) - \dot{x}^0(d^+)) \tag{21} \\
&= C(I - e^{A_1T})^{-1}e^{A_1d}e^{A_1TD}v_s \quad \text{(for buck converter, from (13))} \tag{22} \\
&= C\Lambda(DT)B_1 u \quad \text{(for boost converter, from (19))} \tag{23}
\end{align}$$

where, for large feedback gain, $v_s \approx v_o/D$ for the buck converter and $v_s \approx v_o(1-D)$ for the boost converter. Note that (21), (22), and (23) are exact representations for the S plot, while (15) is an approximate representation, which is generally accurate if the poles are smaller than $\omega_s/10$ as discussed above. Then, from (10), SNB occurs when

$$S(D) = \dot{h}(d) \tag{24}$$

## 5 State-Space Average Analysis

Average analysis can be applied to analyze SNB when the switching condition is determined by the *average* signal values, such as in VMC, or in average CMC. If the switching condition is determined by the *peak* signal values, such as in peak CMC, the average analysis may be inaccurate.

Let $A = DA_1 + (1-D)A_2$, and $B = DB_1 + (1-D)B_2$. In the state-space average model [3], the power stage dynamics is $\dot{x} = Ax + Bu$, and the steady-state solution is

$$-A^{-1}Bu := X \tag{25}$$

Using a hat ˆ to denote small perturbations, (e.g., $\hat{x} = x - X$), the linearized open-loop dynamics [3] is

$$\dot{\hat{x}} = A\hat{x} + ((A_1 - A_2)X + (B_1 - B_2)u)\hat{D} \tag{26}$$

In the closed loop, $\hat{D} = \hat{y}/V_h = C\hat{x}/V_h$. Then, the closed-loop dynamics is

$$\dot{\hat{x}} = (A + \frac{((A_1 - A_2)X + (B_1 - B_2)u)C}{V_h})\hat{x} := \Phi_A \hat{x} \tag{27}$$

When SNB occurs, $\Phi_A$ has an eigenvalue at 0, and $\det[\Phi_A] = 0$. Suppose $\det[A] \neq 0$ then

$$\det[\Phi_A] = \det[A]\det[I + \frac{A^{-1}((A_1 - A_2)X + (B_1 - B_2)u)C}{V_h}] = 0 \tag{28}$$

leading to the following boundary condition

$$V_h + CA^{-1}((A_1 - A_2)X + (B_1 - B_2)u) = 0 \tag{29}$$

The buck converter generally has $A_1 = A_2$ (invertible), $B_{21} == 0_{N\times 1}$, and $B_{12} = B_{22}$, and the boundary condition (29) becomes (equivalent to (18) based on the sampled-data slope-based analysis)

$$V_h + CA^{-1}B_{11}v_s = 0 \tag{30}$$

The boost converter generally has $B_1 = B_2$, and the boundary condition (29) becomes

$$V_h + CA^{-1}(A_1 - A_2)X = 0 \tag{31}$$

Note that the condition (30) is independent of $D$, and is only an indication of the closeness to SNB, while the condition (31) is a function of $D$ through $A = DA_1 + (1-D)A_2$.

# 6 Harmonic Balance Analysis

Applying harmonic balance (HB) analysis to predict *subharmonic oscillation* has been reported in [7, 11]. Here, HB analysis is applied to predict SNB. Consider a buck converter power stage with a control-to-output ($D$-to-$v_o$) transfer function $G_{vd}(s)$. In the converter, there is an ON switch and an OFF switch (sometimes substituted by a diode). Let the voltage across the OFF switch (or the diode) be $v_d$. The waveform of $v_d(t)$ is a square wave with the high voltage at $v_s$ and the low voltage at 0, which can be represented by Fourier series (harmonics)

$$v_d(t) = \sum_{n=-\infty}^{\infty} c_n e^{jn\omega_s t} \quad \text{where} \quad c_n = \frac{v_s}{j2n\pi}(1 - e^{-jn\omega_s d}) \tag{32}$$

The HB analysis can be applied to VMC, or CMC with voltage loop open/closed.

## 6.1 VMC

Let the $v_d$-to-$v_o$ transfer function be $G_v(s)$. One has [3, p. 470]

$$G_v(s) = \frac{G_{vd}(s)}{v_s} = \frac{sR_cC + 1}{LC(1 + \frac{R_c}{R})s^2 + (\frac{L}{R} + R_cC)s + 1} \tag{33}$$

Let the compensator transfer function (from $v_o$ to $-y$ (negative sign due to negative feedback)) be $G_c(s)$. Let the transfer function from $v_d$ to $-y$ be $G(s) = G_c(s)G_v(s) = G_c(s)G_{vd}(s)/v_s$. For designation purpose, $G(s)$ is called a Harmonic Balance (HB) gain, which is proportional to the loop gain $T(s) = G_c(s)G_{vd}(s)/V_h$ by

$$G(s) = \frac{V_h}{v_s}T(s) \tag{34}$$

Then, the signal $y(t)$ at the output of the compensator is

$$\begin{align}
y(t) &= v_r + G_c \star (v_r - v_o) \tag{35}\\
&= (1 + G_c(0))v_r - G_c \star v_o \tag{36}\\
&= (1 + G_c(0))v_r - G_c G_v \star v_d(t) \tag{37}\\
&= (1 + G_c(0))v_r - G \star v_d(t) \tag{38}\\
&= (1 + G_c(0))v_r - \sum_{n=-\infty}^{\infty} c_n e^{jn\omega_s t} G(jn\omega_s) \tag{39}\\
&= (1 + G_c(0))v_r - v_s D G(0) - 2\mathrm{Re}\sum_{n=1}^{\infty} c_n e^{jn\omega_s t} G(jn\omega_s) \tag{40}
\end{align}$$

where Re denotes taking the real part of a complex number, and $\star$ denotes convolution.

The intersection of $y(t)$ and $h(t)$ determines the duty cycle and hence the waveform of $v_d(t)$. By "balancing" the equation $y(t) = h(t)$ (written in Fourier series form) at the switching instants, conditions for existence of periodic solutions and SNB can be derived.

In steady state,

$$y(d) - h(d) = (1 + G_c(0))v_r - \sum_{n=-\infty}^{\infty} c_n e^{jn\omega_s d} G(jn\omega_s) - h(d) = 0 \tag{41}$$

which is an equation in terms of $d = DT$. When SNB occurs, only a *single* solution exists, which means that the curve $y(d) - h(d)$ has a flat slope at the critical value of $d$ when SNB occurs. Differentiate (41) with respective to $d$, one has the following theorem.

**Theorem 2** *Given a closed-loop buck converter with a control-to-output transfer function $G_{vd}(s)$ and a compensator transfer function $G_c(s)$, let the HB gain $G(s) = G_c(s)G_{vd}(s)/v_s$. SNB occurs when*

$$v_s \sum_{n=-\infty}^{\infty} e^{j2n\pi D} G(jn\omega_s) + V_h = 0 \tag{42}$$

The condition (42) can be proved to be equivalent to (7) based on the steady-state analysis or (13) based on the sampled-data slope-based analysis, showing that applying different approaches leads to the same condition. The boundary condition (42) can be expressed in various forms for design purpose. For example, (42) leads to any of the following conditions,

$$v_s G(0) + 2v_s \mathrm{Re}\sum_{n=1}^{\infty} e^{j2n\pi D} G(jn\omega_s) + V_h = 0 \tag{43}$$

$$\mathrm{Re}[\sum_{n=1}^{\infty} e^{j2n\pi D} G(jn\omega_s)] = -\frac{V_h + v_s G(0)}{2v_s} \tag{44}$$

or in terms of $v_s$,

$$v_s = \frac{-V_h}{\sum_{n=-\infty}^{\infty} e^{j2n\pi D}G(jn\omega_s)} \quad (45)$$

$$v_s = \frac{-V_h}{G(0) + 2\text{Re}[\sum_{n=1}^{\infty} e^{j2n\pi D}G(jn\omega_s)]} \quad (46)$$

$$(47)$$

Generally, $G_v(s)$, $G_c(s)$ and thus $G(s)$ are low-pass filters. The summation in (44) can be approximated by the term that involves $G(s)$ with the smallest argument. Therefore, (44) becomes

$$\text{Re}[e^{j2\pi D}G(j\omega_s)] = -\frac{V_h + v_s G(0)}{2v_s} \quad (48)$$

**The "H plot": a Nyquist-like plot in the complex plane.**
Note that the left side of (44) is a function of $D$, $\omega_s$, and the HB gain $G(s)$, where $G(s)$ is further a function of the power stage and compensator parameters. Let

$$H(D) := \sum_{n=1}^{\infty} e^{j2n\pi D}G(jn\omega_s) \quad (49)$$

Then, SNB occurs when

$$\text{Re}[H(D)] = -\frac{V_h + v_s G(0)}{2v_s} \quad (50)$$

For designation purpose, $H(D)$ is called an H plot because it is similar to the Nyquist plot in the complex plane for design purpose. Given a desired range of $D$, one can plot $H(D)$ according to (49) to determine whether SNB occurs in this range of $D$.

### 6.2 CMC with Voltage Loop Open

In CMC with the voltage loop open, the controlled output is the inductor current $i_L$. Let the peak inductor current control signal be $i_c$ (also denoted as $v_r$ in Fig. 1) which controls the peak inductor current. One has $y(t) = i_c(t) - i_L(t)$. Similar to (33), the $v_d$-to-$i_L$ transfer function is [3, p. 470]

$$G_i(s) := \frac{(1 + \frac{R_c}{R})Cs + \frac{1}{R}}{LC(1 + \frac{R_c}{R})s^2 + (\frac{L}{R} + R_c C)s + 1} \quad (51)$$

Since no extra compensator (except the compensating ramp $h(t)$) is added in the current loop, $G_c(s) = 1$, and one has the HB gain $G(s) = G_c(s)G_i(s) = G_i(s)$. Since the expression of $G(s)$ is derived, the rest of analysis is similar as in VMC.

### 6.3 CMC with Voltage Loop Closed

Here, there are two feedback loops. Let $G_c(s)$ be the voltage loop compensator. One has $y(t) = i_c(t) - i_L(t) = G_c \star (v_r - v_o(t)) - i_L(t) = G_c(0)v_r - (G_c \star G_v + G_i) \star v_d(t)$. Then, the HB gain is

$$G(s) = G_c(s)G_v(s) + G_i(s) \quad (52)$$

Since the expression of $G(s)$ is derived, the rest of analysis is similar as in VMC.

# 7 Prediction of SNB Based on Loop Gain

Since the HB gain $G(s)$ is proportional to the loop gain $T(s) = G(s)v_s/V_h$, (42) directly leads to the following theorem

**Theorem 3** *Given a closed-loop buck converter with a loop gain $T(s)$, SNB occurs when*

$$\sum_{n=-\infty}^{\infty} e^{j2n\pi D} T(jn\omega_s) = -1 \tag{53}$$

The condition (53) can be expressed in various forms,

$$T(0) + 2\mathrm{Re}[\sum_{n=1}^{\infty} e^{j2n\pi D} T(jn\omega_s)] = -1 \tag{54}$$

$$\mathrm{Re}[\sum_{n=1}^{\infty} e^{j2n\pi D} T(jn\omega_s)] = -\frac{T(0)+1}{2} \tag{55}$$

The summation in (55) can be approximated by the term that involves $T(s)$ with the smallest argument. Therefore, (44) becomes

$$\mathrm{Re}[e^{j2\pi D} T(j\omega_s)] \approx -\frac{T(0)+1}{2} \tag{56}$$

It should be noted that (56) is only an approximate condition, and the exact condition is (53).

**The "L1 plot" in the real domain.**
Note that the boundary condition (55) is a function of $D$, $\omega_s$, and the loop gain $T(s)$, where $T(s)$ is further a function of $v_s$, $V_h$, the power stage and compensator parameters. Define an L1 plot, which is a *real* function, as

$$L_1(D) := \sum_{k=-\infty}^{\infty} e^{j2k\pi D} T(jk\omega_s) \tag{57}$$

Then, SNB occurs when

$$L_1(D) = -1 \tag{58}$$

**The "L2 plot" in the real domain.**
Since in some situations, $V_h = 0$, such as in CMC with no ramp compensation, $T(s) = G(s)v_s/V_h$ becomes infinite. In such a situation, define an L2 plot, which is also a *real* function, as

$$L_2(D) := \sum_{n=-\infty}^{\infty} e^{j2n\pi D} G(jn\omega_s) \tag{59}$$

Then, from (42), SNB occurs when

$$L_2(D) = -\frac{V_h}{v_s} \tag{60}$$

Since (58), (60) and (13) are exact conditions for occurrence of SNB, they are equivalent, and the L1 and L2 plots have matrix forms

$$L_2(D) = \frac{L_1(D)V_h}{v_s} = -TC(I - e^{A_1 T})^{-1} e^{A_1 d} B_{11} \tag{61}$$

# 8 Buck Converter

This section shows occurrence or non-occurrence of SNB in the buck converter under various control schemes. Boundary conditions for specific schemes are derived if SNB occurs. Similar analysis is applied to the boost converter in the next section.

## 8.1 CMC with Voltage Loop Open

Let the state $x = (i_L, v_C)'$, where $i_L$ is the inductor current, and $v_C$ is the capacitor voltage. Then,

$$A_1 = A_2 = \begin{bmatrix} 0 & \frac{-1}{L} \\ \frac{1}{C} & \frac{-1}{RC} \end{bmatrix}$$

$$B_1 = [B_{11}, B_{12}] = \begin{bmatrix} \frac{1}{L} & 0 \\ 0 & 0 \end{bmatrix}, \quad B_2 = [B_{21}, B_{22}] = \begin{bmatrix} 0 & 0 \\ 0 & 0 \end{bmatrix} \quad (62)$$

$$E_1 = E_2 = \begin{bmatrix} 0 & 1 \end{bmatrix}$$

In CMC with the voltage loop open, a switching occurs when $i_c - i_L(t) = h(t)$. One has $y = i_c - i_L$ and

$$C = [-1, 0] \qquad D = [0, 1]$$
$$CA^{-1}B_{11} = \frac{1}{R}, \; CB_{11} = \frac{-1}{L} \qquad CA_1 B_{11} = 0$$

By simple algebra, the boundary condition (15) leads to

$$D = \frac{1+K}{2} + \frac{L\dot{h}(d)}{v_s} \quad (63)$$

where $K := 2L/RT$ as defined in [3]. Since $D < 1$, then from (63), SNB in the CMC buck converter occurs only when $K < 1$. Also from (63), SNB occurs when the converter is unstable ($D > 1/2$).

The boundary condition (63) can be also derived by the harmonic balance analysis. Based on the facts that for $0 < D < 1$

$$\sum_{k=1}^{\infty} \sin(\frac{2\pi k D}{k}) = \pi(\frac{1}{2} - D)$$

and using (51), the boundary condition (43) becomes

$$v_s G(0) + 2v_s \mathrm{Re} \sum_{n=1}^{\infty} e^{j2n\pi D} G(jn\omega_s) + V_h \approx \frac{v_s}{R} + \frac{v_s T}{L} \sum_{k=1}^{\infty} \sin(\frac{2\pi k D}{\pi k}) + V_h \quad (64)$$

$$= \frac{v_s}{R} + \frac{v_s T}{L}(\frac{1}{2} - D) + V_h = 0 \quad (65)$$

which leads to (63). It is interesting to note that completely different approaches lead to the same condition.

**Example 1.** Consider a CMC buck converter with the following parameters: $v_s = 5$ V, $f_s = 200$ kHz, $R = 5$ Ω, $L = 5$ μH, and $C = 40$ μF. Here, $K = 0.4$ and the converter is at light loading and has both CCM and DCM solutions.

Let $i_c$ be the bifurcation parameter. The bifurcation diagram is shown in Fig. 5. SNB occurs when $i_c = 1.225$, $D = 0.7$, and $v_o = 3.5$. Note that the focus of this paper is on the $T$-periodic solution. Those non-$T$-periodic attractors are not shown in the bifurcation diagram to prevent detraction of the focus on SNB.

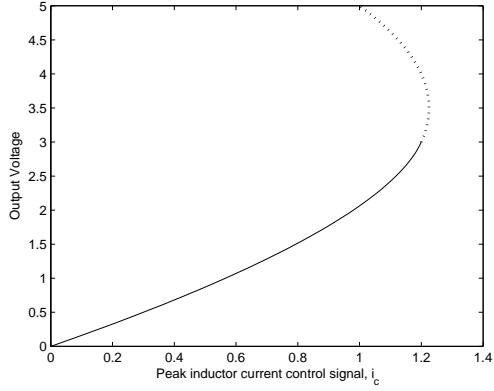

Figure 5: Bifurcation diagram showing SNB at $i_c = 1.225$ and CCM-DCM transition at $i_c = 1.2$, solid line for stable DCM solution, dotted line for unstable CCM solution

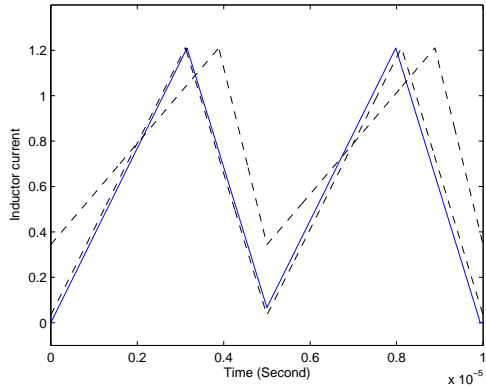

Figure 6: *Coexistence* of two unstable $T$-periodic CCM solutions (dashed line) and a stable $2T$-periodic solution (solid line), $i_c = 1.21$

For $i_c > 1.225$, there is no solution. For $i_c \in (1.2, 1.225)$, there coexist at least three solutions: two unstable (with $D > 1/2$) CCM solutions and a stable attractor. Take $i_c = 1.21$, for example. The two unstable CCM solutions, with duty cycles 0.62 and 0.78, and the stable attractor are shown in Fig. 6. The stable attractor is $2T$-periodic, in CCM for the first period and in DCM (with a tiny duration when the inductor current is zero) for the second period. Note that the three solutions have the same peak inductor current.

As $i_c$ is increased to 1.223, the two unstable CCM solutions move closer, with duty cycles 0.67 and 0.73, as shown in Fig. 7. At $i_c = 1.225$, SNB occurs when the two unstable CCM solutions merge together.

At $i_c = 1.2$, the converter transitions from unstable CCM to stable DCM through a CCM-DCM border-collision bifurcation [12]. For $i_c \in (1, 1.2)$, there exist one stable (with conversion ratio [3] $M < 2/3$) DCM and one unstable CCM (with $D > 1/2$) solutions. For $i_c < 1$, there exists only one stable DCM solution.

The SNB boundary conditions will be derived using various approaches. Here, $K = 0.4$ and $\dot{h}(d) = 0$, then from (63), SNB occurs at $D = 0.7$. The S plot (based on the slope-based analysis), H plot (based on the harmonic balance analysis), L2 (based on the loop gain analysis) are shown in Fig. 8, Fig. 9, and Fig. 10, respectively. They all show that SNB occurs exactly at $D = 0.7$. □

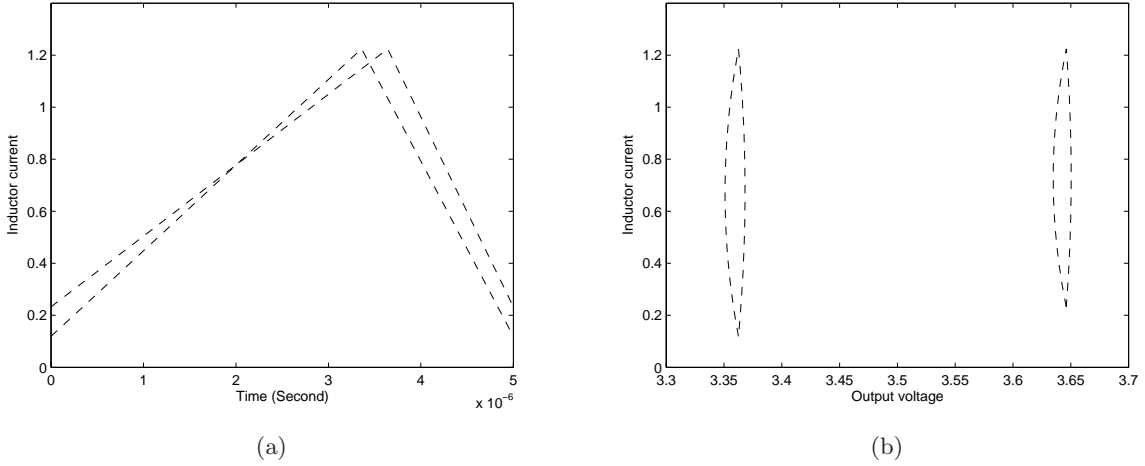

(a)　　　　　　　　　　　　　　　　　(b)

Figure 7: *Coexistence* of two unstable *CCM* solutions, $i_c = 1.223$, (a) time domain, (b) state space

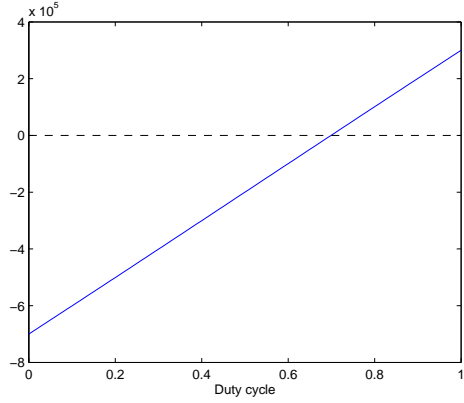

Figure 8: The S plot, $S(D)$

## 8.2  VMC: No SNB

Without loss of generality, let the voltage loop has a proportional feedback gain $k_p$. One has $y = k_p(v_r - v_o)$ and

$$C = [0, -k_p] \qquad\qquad D = [0, k_p]$$
$$CA^{-1}B_{11} = k_p,\ CB_{11} = 0 \qquad CA_1 B_{11} = \frac{-k_p}{LC}$$

The boundary condition (15) becomes

$$-1 + \frac{T^2(1 - 6D + 6D^2)}{12LC} = \frac{V_h}{k_p v_s} \tag{66}$$

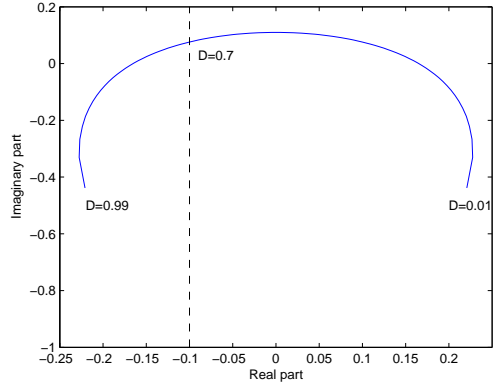

Figure 9: The H plot, $H(D)$

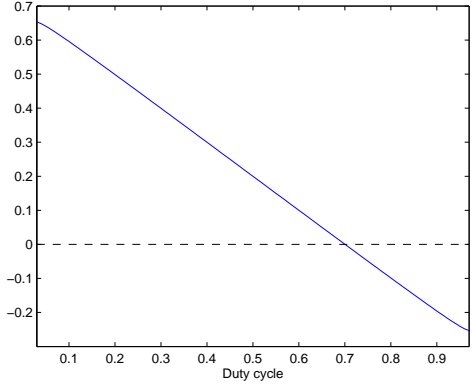

Figure 10: The L2 plot, $L_2(D)$

Generally, $T^2 \ll 12LC$, and the left side of (66) is negative while the right side is positive. The boundary condition is not met, and SNB does not occur in this situation.

## 8.3 Multi-Loop State Feedback

Consider a buck converter under multi-loop state feedback, $y = v_r - k_i i_L - k_v v_C$. One has

$$C = -[k_i, k_v] \qquad\qquad D = [0, 1]$$
$$CA^{-1}B_{11} = \frac{k_i}{R} + k_v,\ CB_{11} = \frac{-k_i}{L} \qquad CA_1 B_{11} = \frac{-k_v}{LC}$$

The boundary condition (15) leads to

$$\frac{v_s k_i}{L}\left(D - \frac{K+1}{2}\right) + \frac{v_s k_v}{T}\left(-1 + \frac{T^2(1 - 6D + 6D^2)}{12LC}\right) = \dot{h}(d) \qquad (67)$$

where the left side of (67) is also an approximate form for the S plot. For $T^2 \ll 12LC$, solving (67) for $D$, SNB occurs when

$$D \approx \frac{K+1}{2} + \frac{L\dot{h}(d)}{v_s k_i} + \frac{Lk_v}{Tk_i} \qquad (68)$$

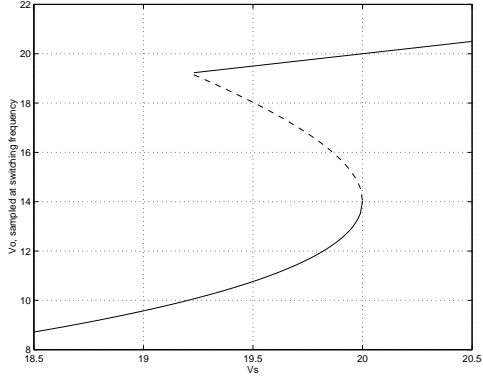

Figure 11: Bifurcation diagram showing stable (solid) and unstable (dashed) solutions. SNB occurs at $v_s = 20$ and $D = 0.7$.

**Example 2.** Consider a buck converter under multi-loop state feedback [1, p. 232]. In [1], a digital control is used, but can be approximated as a multi-loop analog control, with $y(t) = v_r - k_i i_L - k_v v_C$ intersecting with $h(t)$ to determine the duty cycle. The converter parameters are $T = 400$ $\mu$s, $L = 20$ mH, $C = 47$ $\mu$F, $R = 22$ $\Omega$, $k_i = 2.1435$, $k_v = -0.1383$, $v_r = 0.2152$ V, and $V_h = 1$.

Using $v_s$ as the bifurcation parameter, the bifurcation diagram is shown in Fig. 11, showing that SNB occurs at $v_s = 20$ and $D = 0.7$. Generally in SNB, there is a hysteretic loop as shown in Fig. 11. In the figure, the upper solid line is for the operation when the switch is always on (hence $D = 1$), and the dashed line and the lower solid line are for unstable and stable $T$-periodic solutions respectively with duty cycle less than 1. For $v_s$ between 19.25 V and 20 V, there are three solutions: one stable $T$-periodic solution, one unstable $T$-periodic solution, and the third (stable) solution being that the switch is always on. When the converter operates with a periodic solution and $v_s$ is increased a little from 20 V, the output voltage will *jump up* from 14 V to 20 V. Similarly, when the converter operates with $D = 1$ and $v_s$ is decreased a little from 19.25 V, the output voltage will *jump down* from 19.25 V to 10 V. The jumping up and down forms a hysteretic loop. The S plot and its approximate closed form (67) are shown in Fig. 12, both showing that SNB occurs exactly at $D = 0.7$. Also, from (68), using the fact that SNB occurs at $v_s = 20$, then (68) gives $D = 0.71$, which is close to the exact critical value. □

## 9 Boost Converter

### 9.1 VMC

Consider a boost converter under VMC with proportional feedback gain $k_p$. One has $y = k_p(v_r - v_C)$. Let the parasitic resistance associated the inductor be $r$. Then,

$$\begin{aligned}
A_1 &= \begin{bmatrix} \frac{-r}{L} & 0 \\ 0 & \frac{-1}{RC} \end{bmatrix}, \quad A_2 = \begin{bmatrix} \frac{-r}{L} & \frac{-1}{L} \\ \frac{1}{C} & \frac{-1}{RC} \end{bmatrix} \\
B_1 &= B_2 = \begin{bmatrix} \frac{1}{L} & 0 \\ 0 & 0 \end{bmatrix} \\
C &= \begin{bmatrix} 0 & -k_p \end{bmatrix}, \quad D = \begin{bmatrix} 0 & k_p \end{bmatrix} \\
E_1 &= E_2 = \begin{bmatrix} 0 & 1 \end{bmatrix}
\end{aligned} \quad (69)$$

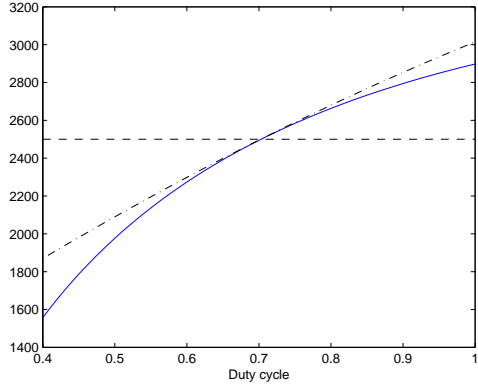

Figure 12: The S plot, $S(D)$ (solid line) and its approximate closed form (67) (dash-dotted line)

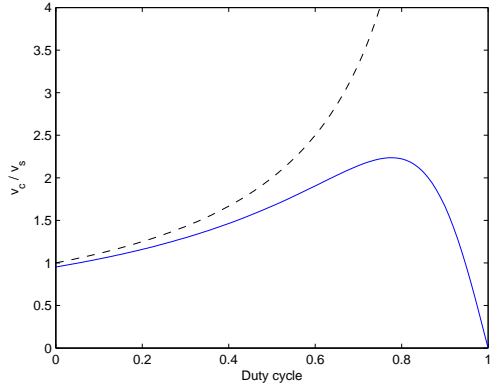

Figure 13: The curve of $v_C(D)/v_s$, for $\rho = 0$ (dashed line) and $\rho = 0.05$ (solid line).

Since the average of $y$ is used in VMC to determine the duty cycle, the average analysis can be applied. Let $\rho = r/R$ and $\kappa = k_p/V_h$. From (25),

$$X = -A^{-1}Bu = \frac{v_s}{\rho + (1-D)^2} \begin{bmatrix} \frac{1}{R} \\ 1-D \end{bmatrix} := \begin{bmatrix} i_L(D) \\ v_C(D) \end{bmatrix} \tag{70}$$

The curve of $v_C(D)/v_s$ is shown in Fig. 13, for $\rho = 0$ and $\rho = 0.05$. For $\rho = 0$, the curve of $v_C(D)/v_s$ increases monotonously as $D$ increases, and one output voltage has only one corresponding duty cycle. For $\rho = 0.05$, the curve of $v_C(D)/v_s$ is $\Lambda$-shaped [3, p. 45]. One output voltage has *two* corresponding duty cycles. The SNB is generally related to coexistence of multiple solutions [1]. It will be shown that SNB occurs only when $\rho > 0$.

Using (69), the boundary condition (31) based on the average analysis leads to

$$(\rho + (1-D)^2)^2 + \kappa v_s((1-D)^2 - \rho)) = 0 \tag{71}$$

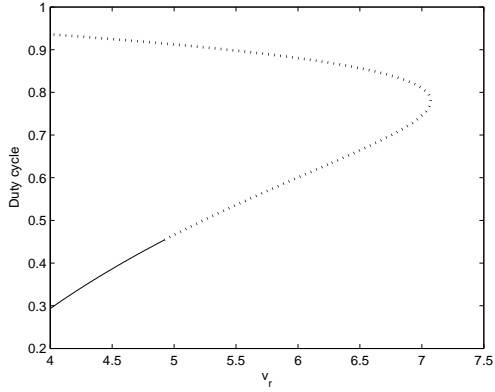

(a)

Figure 14: Bifurcation diagram showing stable (solid) and unstable (dotted) solutions. Hopf bifurcation occurs at $v_r = 4.92$, and SNB occurs at $v_r = 7.1$.

Solving (71) for $D$, then SNB occurs when

$$D = 1 - \sqrt{\sqrt{(2\rho + \frac{\kappa v_s}{4})\kappa v_s} - \rho - \frac{\kappa v_s}{2}} \quad (72)$$

$$\approx 1 - \sqrt{\rho} \text{ (for large } \kappa) \quad (73)$$

If the parasitic inductor resistance $r = 0$ (hence $\rho = 0$), then SNB does not occur for $D < 1$. This shows a dramatic difference if the parasitic inductor resistance $r$ is not considered in the model. If the parasitic resistance is not modeled, one may be misled by the wrong dynamics (no SNB) and steady-state solutions (only one solution), while the actual circuit (with $r > 0$) has SNB and multiple solutions.

**Example 3.** Consider a boost converter under VMC with $k_p = 2$. The converter parameters are $v_s = 3$ V, $V_h = 1$, $f_s = 600$ kHz, $L = 1$ $\mu$H, $C = 100$ $\mu$F, $R = 2$ $\Omega$, and parasitic inductor resistance $r = 0.1$ $\Omega$. One has $\rho = r/R = 0.05$. The bifurcation diagram is shown in Fig. 14. The Hopf bifurcation [13] occurs at $v_r = 4.92$ and SNB occurs at $v_r = 7.1$.

Note that the focus of this paper is on the $T$-periodic solutions. Let $v_r = 7$, for example. The two unstable $T$-periodic solutions are shown in Fig. 15. One solution has $D = 0.74$, and the other has $D = 0.81$, agreed with Fig. 14. Other stable attractors may coexist with the two unstable solutions. In this particular example, the converter is actually bistable for $4 < v_r < 4.92$ and monostable for $v_r > 4.92$. The other stable attractor is $(i_L, v_C) = (v_s/r, 0) = (30, 0)$, and $D = 1$ (one switch is always on) because $y(t) = k_p(v_r - v_C) = k_p v_r > 1 > h(t)$. Such an attractor does not exist if $r = 0$ because it requires $i_L = v_s/r = \infty$. With different converter parameters, the attractor may be quasi-periodic [1], associated with the Hopf bifurcation [13]. Other attractors associated with the border-collision bifurcation [14] may also exist when $y(t)$ of the attractor is out of the bounds of $h(t)$. These non-$T$-periodic attractors are not shown in the bifurcation diagram to prevent detraction of the focus on SNB.

From (72), SNB occurs when $D = 0.78$, the prediction of the SNB agrees with the bifurcation diagram. The S plot (Fig. 16) shows that SNB occurs exactly at $D = 0.78$. □

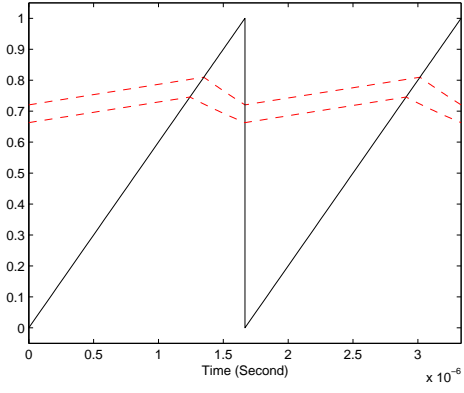

Figure 15: Ramp $h(t)$ (solid line) and two unstable $T$-periodic solutions $y^0(t)$ (dashed line), coexisting with stable attractor $(i_L, v_C) = (30, 0)$

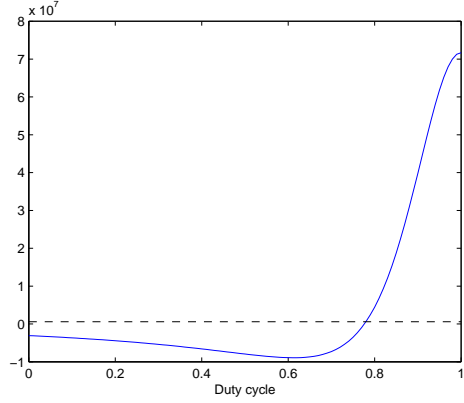

Figure 16: The S plot, $S(D)$

## 9.2 CMC with Voltage Loop Closed

In CMC with the voltage loop closed, the voltage loop output controls the peak inductor current. Without loss of generality, let the voltage loop has a proportional feedback gain $k_p$. In steady state,

$$i_c - i_L = k_p(v_r - v_o) - i_L = h(t) \tag{74}$$

The model is the same as in (69), except that here $C = [-1, -k_p]$.

As in VMC, one can prove that SNB occurs when $\rho > 0$, and SNB does not occur when $\rho = 0$. In steady state, (74) is rearranged as

$$v_r = v_o + \frac{i_L + h(d)}{k_p} \tag{75}$$

For large $k_p$, $(i_L + h(d))/k_p$ can be ignored. Since the curve $v_o$ as a function of $D$ is $\Lambda$-shaped as shown in Fig. 13 if $\rho > 0$, one value of $v_r$ corresponds to two duty cycles and SNB occurs.

In contrast, if $\rho = 0$, $v_o$ increases monotonously as $D$ increases as shown in Fig. 13, one value of $v_r$ corresponds to only one duty cycle and SNB does not occur.

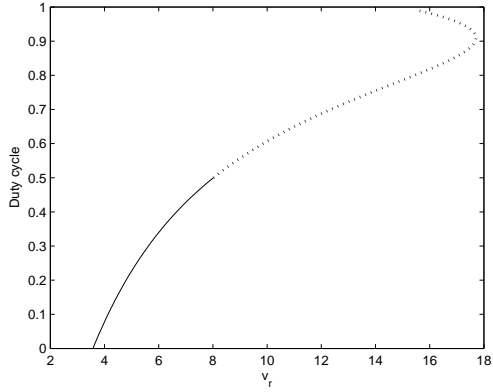

(a)

Figure 17: Bifurcation diagram showing stable (solid) and unstable (dotted) solutions. SNB occurs at $v_r = 17.71$ and $D = 0.91$, and period-doubling bifurcation occurs at $v_r = 8.2$.

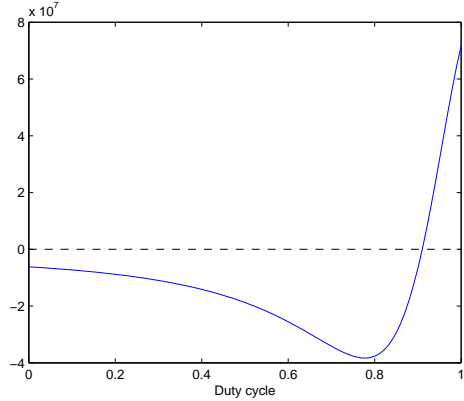

Figure 18: The S plot, $S(D)$

**Example 4.** Consider a boost converter under CMC with no ramp compensation ($V_h = 0$). The voltage loop has a feedback gain $k_p = 2$. The converter parameters are the same as in Example 3.

The bifurcation diagram is shown in Fig. 17, showing that SNB occurs at $v_r = 17.71$ and $D = 0.91$, and the period-doubling bifurcation [13] occurs around $D = 0.5$. The S plot (Fig. 18) shows that SNB occurs exactly at $D = 0.91$. □

### 9.3  CMC with Voltage Loop Open: No SNB

In CMC with the voltage loop open, the current control signal $i_c$ controls the peak inductor current. From (69), the inductor current ripple $\Delta I_L = (v_s - rI_L)DT/L$. The peak inductor current $I_{peak} = I_L + \Delta I_L/2$ increases monotonously as $D$ increases. Therefore, given a value of current control signal $i_c$, only one solution exists and SNB does not occur.

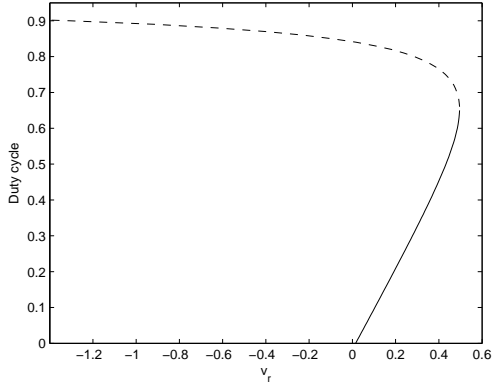

Figure 19: Bifurcation diagram showing stable (solid) and unstable (dashed) solutions. Saddle-node bifurcation occurs at $v_r = 0.496$ and $D = 0.65$.

### 9.4 Multi-Loop State Feedback

Consider a boost converter under multi-loop state feedback, $y = v_r - k_i i_L - k_v v_C$. The model is the same as in (69), except that here $C = -[k_i, k_v]$ and $D = [0, 1]$. The sampled-data analysis is similar to the case for the boost converter under VMC, omitted here to save space. The average analysis can be also applied to obtain an approximate boundary condition. From (31), the boundary condition leads to

$$\frac{V_h}{v_s}(1-D)^2 + \frac{2k_i}{R(1-D)} + k_v = 0 \qquad (76)$$

Since $D < 1$, SNB does not occur if both $k_i$ and $k_v$ are positive because the condition (76) is not met. SNB occurs only when at least one of $k_i$ and $k_v$ is negative.

**Example 5.** Consider a boost converter under multi-loop state feedback [15, p. 90]. The converter parameters are $T = 2$ $\mu$s, $V_s = 4$ V, $L = 5.24$ $\mu$H, $C = 0.2$ $\mu$F, $R = 16$ $\Omega$, $k_i = -0.1$, $k_v = 0.01$, $v_r = 0.48$, and $V_h = 1$. This converter has been analyzed with different views. In [15], this converter is used to illustrate the inaccuracy of the average model about the converter stability. It is later proved in [7, p. 75] that the average model is actually accurate about *local* stability, but the converter is not globally stable. It is then shown in [16] that two solutions (one stable and the other unstable) coexist, but without linking multiple solutions with SNB.

The bifurcation diagram is shown in Fig. 19, showing that SNB occurs at $v_r = 0.496$ and $D = 0.65$. For $v_r = 0.48$, there are actually three solutions: one stable $T$-periodic solution, one unstable $T$-periodic solution, and the third (stable) solution being that one switch is always on (which does not exist in the real circuit because infinite inductor current is required). The S plot (Fig. 20) shows that SNB occurs exactly at $D = 0.65$. Applying the average analysis can also give approximate results. Solving (76) gives $D = 0.685$, which is close to the exact critical value. □

## 10 Conclusion

A unified model of VMC/CMC is proposed to predict SNB. In the unified model, confirmed with simulations and the exact sampled-data analysis [7, 8], SNB boundary conditions for VMC/CMC are derived and they have similar forms for VMC/CMC. The boundary conditions are expressed in terms of signal slopes, or other converter parameters. The sloped-based boundary conditions derived are *exact*, and can be further simplified in various approximate closed forms for design purpose. The sloped-based boundary conditions are applied to analyze buck and boost converters under various VMC/CMC control schemes. Since all the boundary conditions are derived in closed forms, one can see the effects of various converter parameters on the occurrence of SNB.

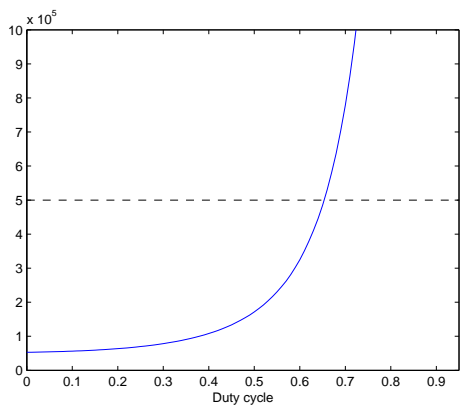

Figure 20: The S plot, $S(D)$

Many approaches can be applied to analyze SNB, including

1. simulation to plot the bifurcation diagram
2. average analysis
3. sampled-data small-signal analysis using the Jacobian matrix to determine the pole location
4. steady-state analysis to determine whether multiple solutions exist
5. sampled-data slope-based analysis and using the S plot
6. harmonic balance analysis and using the H plot
7. loop gain analysis and using the L1/L2 plot

The first three approaches are known, but no closed form boundary conditions have been derived in the past research. The last four approaches are first proposed in this paper to derive the boundary conditions to the author's knowledge. Applying different approaches lead to the same boundary conditions, providing convincing evidences about the accuracy of the derived conditions. Each approach has its own merits, and complement the other approaches. For example, the slope-based analysis analyzes the converter in the time domain and expresses the boundary conditions in matrix forms. The harmonic balance analysis analyzes the converter in the frequency domain and expresses the boundary conditions in terms of harmonics of the switching frequency. Since most power stage and compensator transfer functions are low-pass filters, the harmonic balance analysis is particularly useful to analyze the converter with poles close to the switching frequency.

Similar to the Nyquist plot which predicts stability for the continuous-time system, many new plots are proposed to predict SNB. The S plot is in matrix form, while the other plots are in terms of harmonics. The H plot is in the complex plane, while the other plots are in the real domain. All of these plots are exact and can accurately predict SNB. These plots also have simplified forms for design purpose.

Coexistence of multiple steady-state solutions is generally related to SNB, and steady-state analysis can predict SNB. Generally, SNB occurs when two solutions (with different duty cycles) coexist corresponding to the same value of a bifurcation parameter. Therefore, the shape (monotonously increasing or $\Lambda$-shaped) of a controlled signal in steady state as a function of duty cycle determines whether SNB occurs. For example, in the CMC buck converter, the peak inductor current is controlled and its steady-state curve as a function of duty cycle is $\Lambda$-shaped. Two steady-state solutions of different duty cycles have the same value of peak inductor current, and SNB occurs in the CMC buck converter. For the boost converter, the steady-state output voltage as a function of duty cycle is $\Lambda$-shaped (see Fig. 13) if the parasitic inductor resistance is modeled. Therefore, SNB occurs in the boost converter if the output voltage is controlled, such as in VMC, or in CMC with the voltage loop closed. However, if the parasitic inductor resistance is zero, the steady-state output voltage monotonously increases as the duty cycle increases (see Fig. 13), SNB does not occur in VMC, or in CMC with the voltage loop closed if the parasitic inductor resistance is zero. Similarly, in the boost converter, the steady-state inductor current monotonously increases as the duty cycle increases, and SNB does not occur if the inductor current is controlled, such as in CMC with the voltage loop open.

# References

[1] C.-C. Fang and E. H. Abed, "Saddle-node bifurcation and Neimark bifurcation in PWM DC-DC converters," in *Nonlinear Phenomena in Power Electronics: Bifurcations, Chaos, Control, and Applications*, S. Banerjee and G. C. Verghese, Eds. New York: Wiley, 2001, pp. 229–240.

[2] R. B. Ridley, "A new continuous-time model for current-mode control with constant on-time, constant off-time, and discontinuous conduction mode," in *Proc. IEEE PESC*, 1990, pp. 382–389.

[3] R. W. Erickson and D. Maksimovic, *Fundamentals of Power Electronics*, 2nd ed. Berlin, Germany: Springer, 2001.

[4] S. Pavljasevic and D. Maksimovic, "Using a discrete-time model for large-signal analysis of a current-programmed boost converter," in *Proc. IEEE PESC*, 1991, pp. 715–721.

[5] Y. Ma, C. K. Tse, T. Kousaka, and H. Kawakami, "Connecting border collision with saddle-node bifurcation in switched dynamical systems," *IEEE Transactions on Circuits and Systems II: Express Briefs*, vol. 52, no. 9, pp. 581–585, 2005.

[6] C.-C. Fang, "Unified discrete-time modeling of buck converter in discontinuous mode," *IEEE Trans. Power Electron.*, vol. 26, no. 8, pp. 2335–2342, 2011.

[7] ——, "Sampled-data analysis and control of DC-DC switching converters," Ph.D. dissertation, Dept. of Elect. Eng., Univ. of Maryland, College Park, 1997, available: http://www.lib.umd.edu/drum/, also published by UMI Dissertation Publishing in 1997.

[8] C.-C. Fang and E. H. Abed, "Sampled-data modeling and analysis of closed-loop PWM DC-DC converters," in *Proc. IEEE ISCAS*, vol. 5, 1999, pp. 110–115.

[9] ——, "Limit cycle stabilization in PWM DC-DC converters," in *IEEE Conference on Decision and Control*, 1998, pp. 3046–3051.

[10] Y. A. Kuznetsov, *Elements of Applied Bifurcation Theory*. New York: Springer-Verlag, 1995.

[11] C.-C. Fang and E. H. Abed, "Harmonic balance analysis and control of period doubling bifurcation in buck converters," in *Proc. IEEE ISCAS*, vol. 3, May 2001, pp. 209–212.

[12] S. Parui and S. Banerjee, "Bifurcations due to transition from continuous conduction mode to discontinuous conduction mode in the boost converter," *IEEE Trans. on Circuits and Systems-I: Fundamental Theory and Applications*, vol. 50, no. 11, pp. 1464–1469, 2003.

[13] C. K. Tse and M. Di Bernardo, "Complex behavior in switching power converters," *Proceedings of the IEEE*, vol. 90, no. 5, pp. 768–781, 2002.

[14] G. Yuan, S. Banerjee, E. Ott, and J. A. Yorke, "Border-collision bifurcations in the buck converter," *IEEE Transactions on Circuits and Systems-I: Fundamental Theory and Applications*, vol. 45, no. 7, pp. 707–716, 1998.

[15] B. Lehman and R. M. Bass, "Switching frequency dependent averaged models for PWM DC-DC converters," *IEEE Trans. Power Electron.*, vol. 11, no. 1, pp. 89–98, 1996.

[16] J. W. van der Woude, W. L. de Koning, and Y. Fuad, "On the periodic behavior of PWM DC-DC converters," *IEEE Trans. Power Electron.*, vol. 17, no. 4, pp. 585–595, 2002.